\documentclass[]{an}
\usepackage{times}
\usepackage{mathptm} 
\usepackage{natbib}
\usepackage[dvips]{graphicx}

\Volume{00}              
\Year{0000}              
\Month{00}               
\Pagespan{000}{000}      

\newcommand{\as}{$^{\prime\prime}\!\!.$}
\newcommand{\ha}{H$\alpha$ }
\newcommand{\hb}{H$\beta$ }
\newcommand{\oiii}{[O\,III] }
\newcommand{\nii}{[N\,II] }
\begin{document}
\lhead[\thepage]{K.\ Jahnke et al.: Integral field spectroscopy of QSO host galaxies}
\rhead[Astron. Nachr./AN~{\bf XXX} (200X) X]{\thepage}
\headnote{Astron. Nachr./AN {\bf 32X} (200X) X, XXX--XXX}

\title{Integral field spectroscopy of QSO host galaxies}

\author{K.~Jahnke\inst{1}
  \and L.~Wisotzki\inst{1,2}
  \and S.~F.~S\'anchez\inst{1}
  \and L.~Christensen\inst{1}
  \and T.~Becker\inst{1}
  \and A.~Kelz\inst{1} 
  \and M.~M. Roth\inst{1}
}
\institute{Astrophysikalisches Institut Potsdam, An der Sternwarte 16, 14482 Potsdam, Germany
\and  Potsdam University, Am Neuen Palais 10, 14469 Potsdam, Germany
}
\correspondence{http://www.aip.de/$\mathrm\sim$jahnke/}
\date{Received / Accepted}

\abstract{We describe a project to study the state of the ISM in
$\sim$20 low redshift ($z<0.3$) QSO host galaxies observed with the
PMAS integral field spectrograph. We describe method developement to
access the stellar and gas component of the spectrum without the
strong nuclear emission to access the host galaxy properties also in
the central region. It shows that integral field spectroscopy promises
to be very efficient to study the gas distribution and its velocity
field, and also spatially resolved stellar population in the host
galaxies also of luminous AGN.
}

\maketitle

\section{Introduction}
\label{intro}

The vast majority of QSO host galaxy studies uses imaging techniques
to investigate photometry, the morphological composition, signs for
tidal interaction and star formation, companions or clustering. With
imaging it is possible to constrain the types of the host galaxies
which turned out to be dependent on the luminosity of the nucleus
\citep[e.g.][]{mcle95a}, or vice versa. Using imaging photometry we
recently found signs for bluer colours in elliptical host galaxies
\citep{jahn03e} and obtained the first luminosity function of host
galaxies \citep{kuhl01,kuhl03}. Imaging studies showed also e.g.\ that
interaction seems to play a major role in a significant fraction of
intermediately luminous host galaxies, but clearly not in some parts
of the high luminosity regime \citep{dunl03}.

Imaging was and is successful from the closest AGN up to high
redshifts and by now some standard analysis methods have emerged, like
two-dimensional modelling of the surface brightness distribution of
nucleus and host galaxy. This allows to determine and remove the
nuclear contribution to access the light of the host alone -- being
the fundamental problem of host galaxy studies.

Spectroscopy in comparison is used only in a few studies. The
application of spectroscopy is even stronger constrained than imaging
by the presence of the bright nucleus. Studies are in most cases
restricted to the outer part of the host galaxy, away from the light
emitted by the nucleus -- so the inner parts of host galaxies were
unaccessible. Placing a slit across the nucleus, the obvious choice
for an investigation of the inner part of the host, would as a
disadvantage have integrated most of the point like nuclear flux, but
only very little of the host galaxies and thus leading to an
unfavorable contrast. Only very recently, newly developed techniques
involving deconvolution \citep{cour02b} and again spatial modelling of
the light distribution \citep{jahn02} made on-nuclear spectroscopy
feasible, with some success.

We want to describe here an approach using integral field units for
efficient spectroscopy of quasar host galaxies, combining imaging and
spectroscopy in one observation. While the analysis is just fully
beginning, we want to present our goals, the data obtained and, in
Section~\ref{methods}, a technique and an example for the removal of
the nuclear light, required for studying the inner parts of the host
galaxies.

\section{Main science drivers}
\label{project}

The study we describe in the following was triggered by a number of
independent questions, each connecting the active nucleus and the
surrounding host galaxy:

\begin{itemize}
\item It is still unclear how AGN are fueled. The widespread idea of
an environmental triggering event e.g.\ by major or minor mergers is,
however, still not proven. Tidal events should leave traces in the
morphology of the host galaxies. Postulating this as the sole
mechanism for activity is challenged by existence of seemingly
completely symmetric and undisturbed host galaxies.
\item How does the the strong UV and optical nuclear radiation
influence the conditions in the host galaxy? Feedback of
high-luminosity radiation of the nucleus into the host will ionise
existing ISM.
\item Besides ionisation by the nucleus, is there evidence for
star-formation, maybe created by interaction?
\item If some of the UV radiation gets absorbed in the ISM of the
host, does any of it escape? At earlier epochs QSOs contributed a
large fraction of the metagalactic UV background. Low redshift AGN
provide the ideal laboratory to study the escape mechanisms in detail.
\item While the unified model of AGN \citep{urry95} predicts an
absorbing dust torus around the central engine to create a
line-of-sight dependent family of type~1 and 2 AGN, there is no a
priori requirement for an alignment of such a torus with any component
of the host galaxy. Do low redshift disk-type QSO host galaxies show
ionisation cones produced by the `shadow' of such a torus, thus e.g.\
pointing to a non-alignment of torus and a stellar disk?
\item Are there dynamical traces for merger events in the host judging
from dynamical information extracted from emission line shifts?
\end{itemize}

The processes behind these questions all leave traces in the ISM of
the host galaxy in the shape of the two-dimensional distribution of
ionisation states and gas velocity fields. For efficient observation
of these properties IFS promises to be less time consuming than
combinations of long slit spectroscopy or narrow band imaging.

\section{Sample and observations}
We selected a sample of approximately 40 low redshift ($z<0.3$)
quasars from the Palomar-Green \citep[PG,][]{schm83} and Hamburg/ESO
\citep{wiso00} surveys with $V$ band magnitudes in the range $V=
14-17$. The PG objects have previously been studied by others using
broad band imaging \citep{mcle95a}.

PMAS, the Potsdam Multiaperture Photospectrometer, available as a
common user instrument at the Calar Alto 3.5m telescope, showed to be
well suited for this project. PMAS is a fibre fed integral field
spectrograph with a 16$\times$16 lenslet array and covering a FOV of
8$''\times ''$8 with a spaxel size of 0\as5 \citep[described in detail
in][]{roth00}. We observed about half of this sample in two campaigns
in 09/02 and 05/03, using the V300 grisms giving a spectral resolution
of $\sim 6$\AA. The grating was rotated to an angle to observe the
H$\beta$/\oiii and H$\alpha$/\nii regions simultaneously, and we used
integration times on target of typically 1--2~h, split into several
individual integrations for cosmic ray removal. During the campaigns
the seeing was strongly varying between 0\as8 and 1\as7.

The data extraction and reduction of the spectra was done with
P3D\_online, an IDL based software package created at the AIP for
reduction of 3d data, from PMAS in particular \citep{beck01}. Analysis
of the data is in the starting stage.

\begin{figure}
\centering \resizebox{\hsize}{!}{\includegraphics{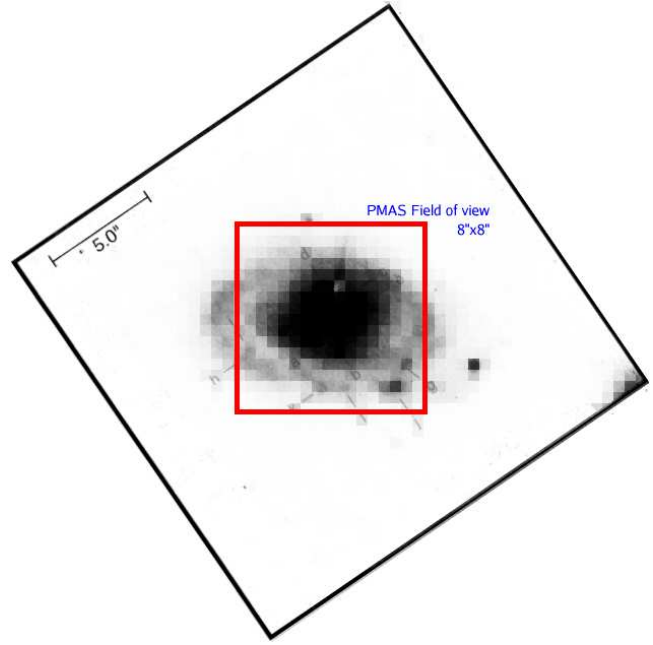}}
\caption{
PMAS resolution and field-of-view of the PMAS spectrograph on the
example of PG\,0052+251 ($z=0.155$). The HST image is taken from
\citep{bahc96}. The host galaxy effectively fills the FOV. 
}\label{fig:PGexample}
\end{figure}

\section{3d analysis methods}
\label{methods}

3d spectroscopy potentially allows the synergy of advantages of
diagnostics from imaging and spectroscopy. It allows to switch between
spectral and imaging representation depending on the application as
well as creating completely new representations like 2d velocity
fields from a single observation. However, for this to work new
analysis techniques are required. As mentioned, the central challenge
when investigating QSO host galaxies compared to inactive galaxies is
the active nucleus. Being very luminous -- up to several times the
host galaxy luminosity -- it appears as a bright point source in the
centre of the host. With a finite PSF size it outshines some portion
of the host galaxy, requiring to remove a large fraction of the
nuclear contribution before the underlying stellar and nebular
contributions become accessible to analysis. We want to sketch in the
following two ideas for nucleus removal in the context of IFS.

\subsection{Line emission vs.\ continuum}
Today the fields of view of IFU instruments are relatively small, for
imaging type applications $<60''$. Effectively this prevents the
simultaneous observation of a nearby star to determine from its
spatial light distribution the PSF at the moment of
observation. Taking PG\,0052+251 in Fig.~\ref{fig:PGexample} as an
example\footnote{Full reduction and results on PG\,0052+251 as well as
the other objects in the sample will be presented in an upcoming
paper.}, the host already largely fills the FOV.


\begin{figure}
\centering \resizebox{\hsize}{!}{\includegraphics[clip,angle=0,width=15cm]{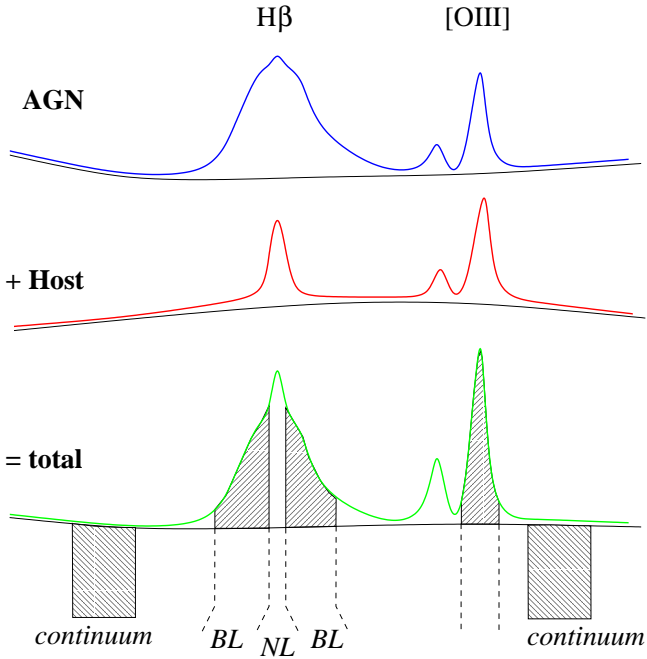}}
\caption{
Schematic composition of a QSO spectrum. Narrow band image slices can
be extracted and with removal of an appropriate continuum slice pure
line flux images are constructed in broad \hb (pure nucleus) and \oiii
(mixed nucleus NLR and host ISM).
}\label{fig:psfsketch}
\end{figure}

\begin{figure*}
\centering \resizebox{\hsize}{!}{\includegraphics[clip,width=5cm,angle=180]{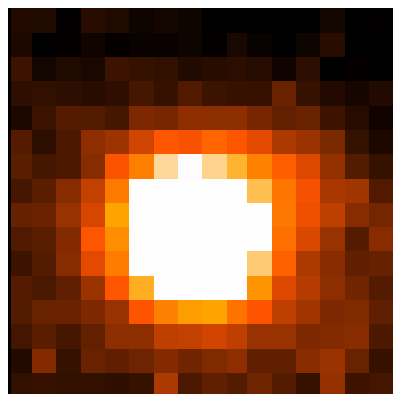}\hspace{5mm}\includegraphics[clip,width=5cm,angle=180]{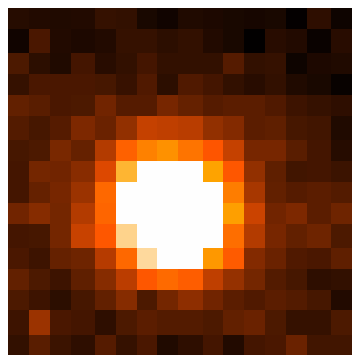}\hspace{5mm}\includegraphics[clip,width=5cm,angle=180]{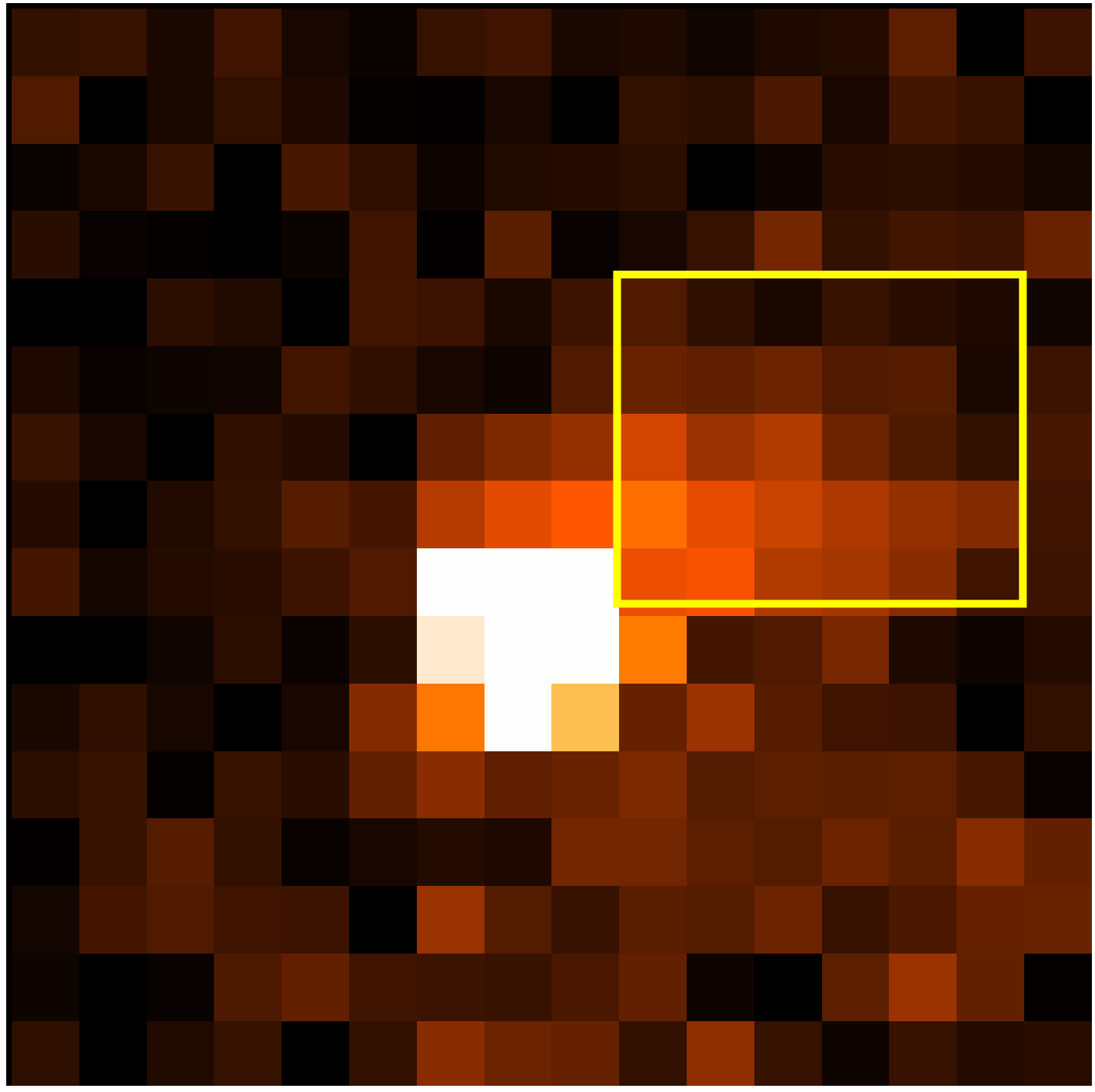}}
\caption{
Extracted image slices of the \oiii line incl.\ continuum (left), a
continuum region next to \oiii and difference between those two images
(right) showing the spatial structure in the \oiii emission. The
square marks the region where the spectrum in Fig.~\ref{fig:bothspecs}
was extracted from. There are no signs for stellar absorption lines
visible so the spectrum is dominated by the nuclear continuum and
lines.
}\label{fig:OIIIframes}
\end{figure*}

\begin{figure}
\centering \resizebox{\hsize}{!}{\includegraphics[clip,angle=0]{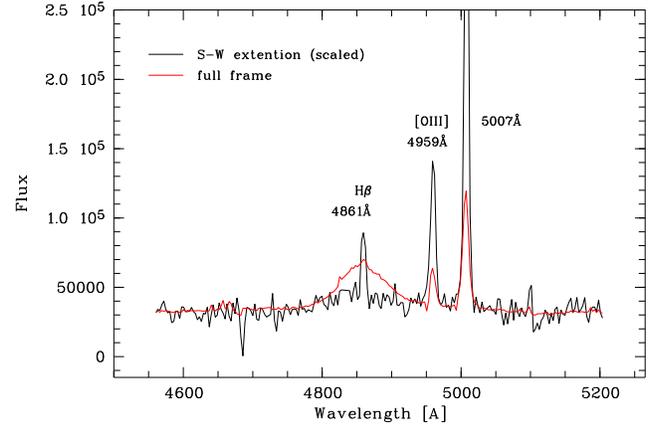}}
\caption{
Total spectrum (lighter red line) and spectrum from extended \oiii
emission region in Fig.~\ref{fig:OIIIframes}. While \hb is dominated
by broad line emission in the full spectrum, only narrow emission is
visible in the extended region thus this is free from nuclear
emission.
}\label{fig:bothspecs}
\end{figure}

Without the PSF available, it is still possible to extract information
on ionisation structure in the host by creating pure line emission
image slices. The total spectrum of the QSO is a mix of almost pure
nuclear continuum emission and lines from nucleus and the ISM of the
host (red line in Fig.~\ref{fig:bothspecs}). Using PMAS as a tunable
narrow band filter, we coadded the monochromatic slices of the \oiii
region and a nearby line-free continuum region as illustrated in
Fig.~\ref{fig:psfsketch} to produce a pure \oiii line image
(Fig.~\ref{fig:OIIIframes}). Besides the nuclear component it shows an
assymetric extention. Coadding the spectra in this region shows no
more broad component in the \hb region, only a narrow component is
left over (Fig.~\ref{fig:bothspecs}). Thus we have no contamination
from the nuclear light. Measuring the emission line ratio
$\log(\mathrm{[OIII]/\mathrm{H}\beta})$ we find a value of 1.0,
typical for excitation of oxygen by an AGN type power-law spectrum and
not by star formation.

\subsection{PSF determination from QSO science frame}
\label{psf}

Even though in this way we can receive already some information about
emission line structure, quantitative studies of the host galaxy or
ISM are not possible by simple continuum removal. In case of PMAS
there exist two possibilities to construct a PSF for removing the
complete nuclear contribution. The first -- to be investigated in the
near future -- is the mapping of the PMAS-internal autoguider camera
PSF to the spectrograph. This will allow the determination of a high
precision PSF if sufficient numbers of point sources have been
observed to calibrate the system for different atmospheric and
instrumental conditions.

\begin{figure*}
\centering \resizebox{\hsize}{!}{\includegraphics[width=9.5cm]{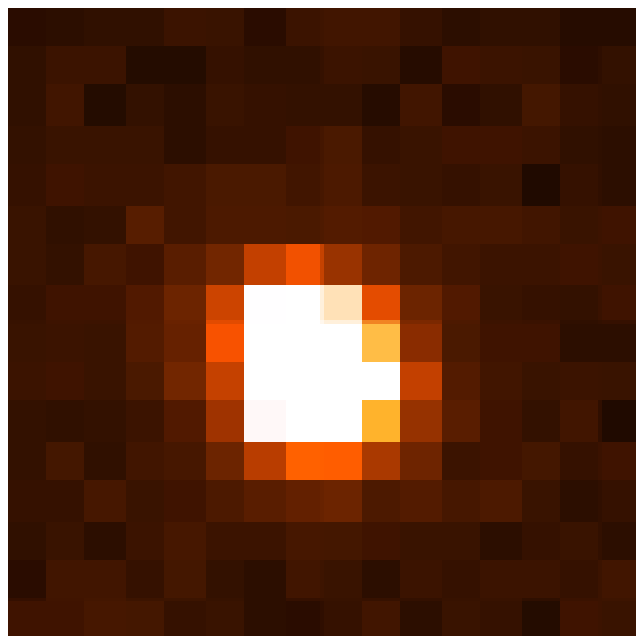}\hspace{7mm}\includegraphics{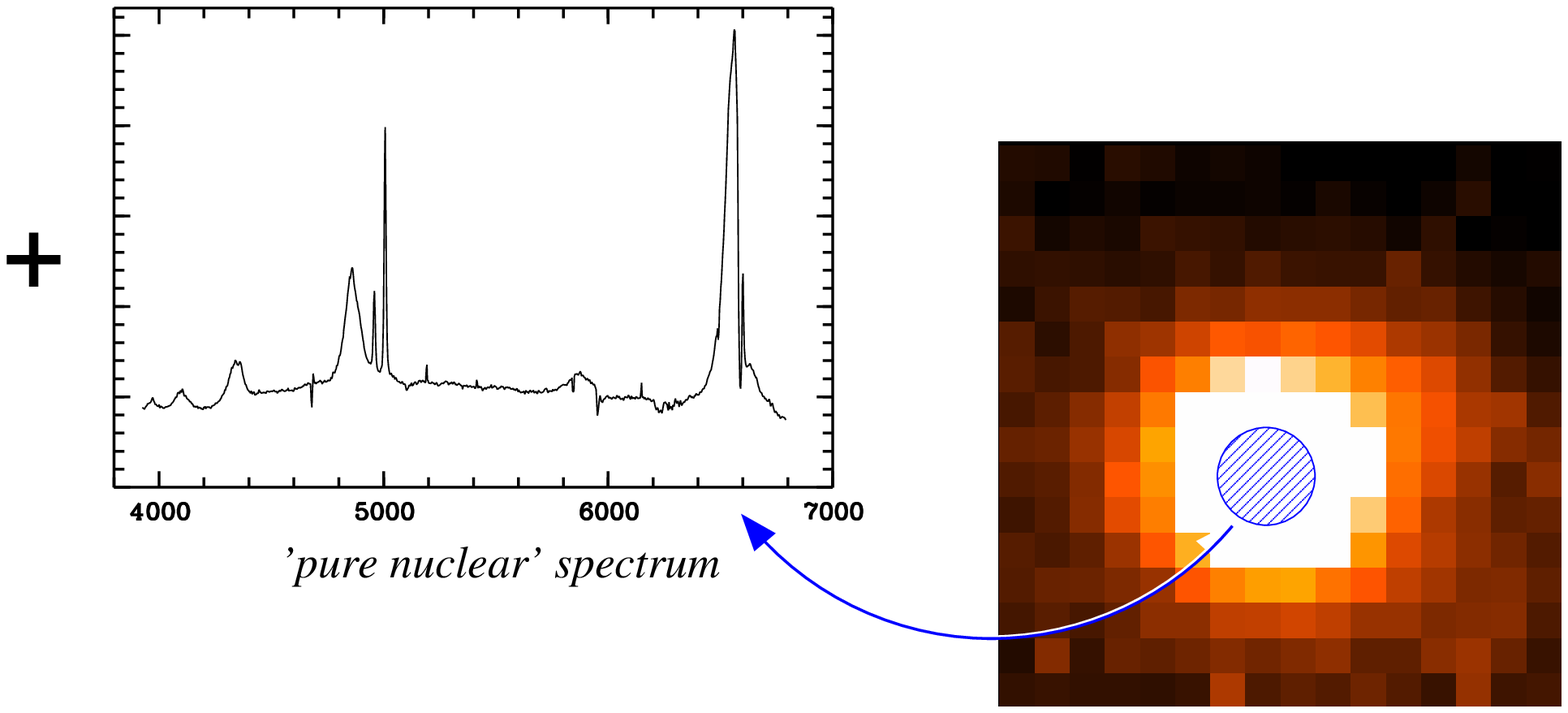}
}
\caption{
2d PSF frame (left) extracted from \hb broad line component of a QSO
frame. The FWHM is 0\as9, thus slightly undersampling the
spectrograph's PSF with 0\as5 spaxels. Combined with a 1d `purely
nuclear' spectrum a 3d PSF can be reconstructed.
}\label{fig:psfextraction}
\end{figure*}

The second is only possible for QSOs but is independent of the AG
camera: a single component in the spectrum of a QSO is solely due to
the active nucleus: the broad line emission of e.g.\ the Balmer
lines. In our case \ha and \hb have a core of a mix of narrow line and
broad line emission, but the wings are purely nuclear emission as
sketched in Fig.~\ref{fig:psfsketch}. Adding up the corresponding
image slices and subtracting an appropriate background frame removes
all of the contained host and nuclear continuum emission, resulting in
a pure PSF (left panel of Fig.~\ref{fig:psfextraction}).

To recreate a 3d PSF from this 2d image can be done by creating an
image cube from this single frame, scaling its flux with an
approximated `purely nuclear' 1d spectrum that can be obtained by
extracting a spectrum from the centermost few pixels in each slice,
where the nucleus strongly dominates (Fig.~\ref{fig:psfextraction},
right panel). Strictly speaking the 2d PSF is only valid for the
spectral position from where it was taken, in our case the \hb
region. When recreating the 3d PSF in this way both change in center
from atmospheric differential diffracion as well as the wavelength
dependent width of the PSF should be taken into account. This can be
done e.g.\ by interpolating between the PSFs from different broad
lines or a computational broadening according to a theoretical seeing
model.

\begin{figure}
\centering
\includegraphics[bb=101 101 300 300,clip,angle=180,width=5.5cm]{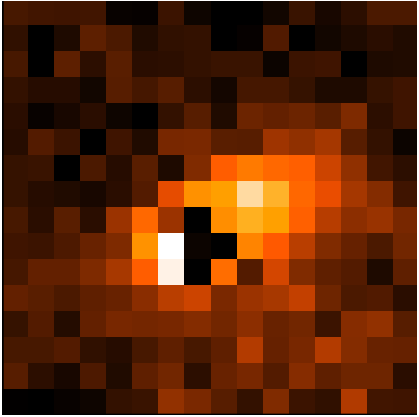}
\caption{
PSF subtracted \oiii image. In the center some positive and negative
residuals of the subtraction are visible.
}\label{fig:OIII_subtracted}
\end{figure}

When subtracting this PSF from the \oiii frame in
Fig.~\ref{fig:OIIIframes}, we receive the image shown in
Fig.~\ref{fig:OIII_subtracted}. The frame shows adjacent positive and
negative residuals of the order of $\sim5$\,\%, due to a slight
mismatch in centering. The very central parts of PSF subtracted frames
will remain with residual contamination from the strong PSF from small
PSF errors and show a much higher noise. Visible is extended \oiii
emission apart from the structure seen before, to the right and
bottom. This is the trace of extended ISM gas, ionised by the
nucleus. This image can now be used to measure emission line ratios
and shifts to create maps of the ionisation source and the velocity
structure in the host galaxy.

\section{Summary}
\label{summary}
It is clear that 3d spectroscopy requires new methods for analysis
also for the application to the study of AGN host galaxies. As we
described it is possible to remove the light of the active nucleus and
to extract information on the host galaxy alone, allowing the use of
techniques as for inactive galaxies. In this way 3d spectroscopy
becomes extremely powerful for the study of QSO host
galaxies. Spatially resolved spectral information can be the key in
the search for traces in the dynamic and ionisation state of the ISM
in host galaxies, to yield information on the role of merger and
interaction in the fuelling process.

\end{document}